\pdfoutput=1

\documentclass[twocolumn,superscriptaddress,floatfix]{revtex4}

\usepackage{amsmath}
\usepackage{amssymb}

\usepackage{graphicx}

\usepackage{color} 

\usepackage[labelfont=bf,labelsep=period,justification=raggedright]{caption}

\makeatletter
\renewcommand{\@biblabel}[1]{\quad#1.}
\makeatother

\newcommand{\dadi}{$\partial$a$\partial$i}

\begin{document}

\title{Inferring the joint demographic history of multiple populations from multidimensional SNP frequency data}

\author{Ryan N. Gutenkunst}
\affiliation{Theoretical Biology and Biophysics \& Center for Nonlinear Studies, Los Alamos National Laboratory, Los Alamos, New Mexico, USA}

\author{Ryan D. Hernandez}
\affiliation{Human Genetics, University of Chicago, Chicago, Illinois, USA}

\author{Scott H. Williamson}
\affiliation{Biological Statistics and Computational Biology, Cornell University, Ithaca, New York, USA}
\author{Carlos D. Bustamante}
\affiliation{Biological Statistics and Computational Biology, Cornell University, Ithaca, New York, USA}

\begin{abstract}
Demographic models built from genetic data play important roles in illuminating prehistorical events and serving as null models in genome scans for selection.
We introduce an inference method based on the joint frequency spectrum of genetic variants within and between populations.
For candidate models we numerically compute the expected spectrum using a diffusion approximation to the one-locus two-allele Wright-Fisher process, involving up to three simultaneous populations.
Our approach is a composite likelihood scheme, since linkage between neutral loci alters the variance but not the expectation of the frequency spectrum.
We thus use bootstraps incorporating linkage to estimate uncertainties for parameters and significance values for hypothesis tests.
Our method can also incorporate selection on single sites, predicting the joint distribution of selected alleles among populations experiencing a bevy of evolutionary forces, including expansions, contractions, migrations, and admixture.

We model human expansion out of Africa and the settlement of the New World, using 5 Mb of noncoding DNA resequenced in 68 individuals from 4 populations (YRI, CHB, CEU, and MXL) by the Environmental Genome Project. 
We infer divergence between West African and Eurasian populations 140 thousand years ago (95\% confidence interval: 40 -- 270 kya).  
This is earlier than other genetic studies, in part because we incorporate migration.  
We estimate the European (CEU) and East Asian (CHB) divergence time to be 23 kya (95\% c.i.: 17 -- 43 kya), long after archeological evidence places modern humans in Europe.
Finally, we estimate divergence between East Asians (CHB) and Mexican-Americans (MXL) of 22 kya (95\% c.i.: 16.3 -- 26.9 kya), and our analysis yields no evidence for subsequent migration.  
Furthermore, combining our demographic model with a previously estimated distribution of selective effects among newly arising amino acid mutations accurately predicts the frequency spectrum of nonsynonymous variants across three continental populations (YRI, CHB, CEU).

\vspace{0.25in}

\noindent {Abbreviations: AFS, allele frequency spectrum; CEU, Utah residents of Northern and Western European ancestry; CHB, Han Chinese from Beijing, China; EGP, Environmental Genome Project; kya, thousands of years ago; LD, linkage disequilibrium; MXL, Los Angeles residents of Mexican ancestry;  SNP, single nucleotide polymorphism; YRI, Yoruba from Ibadan, Nigeria}

\end{abstract}

\maketitle


\section{Introduction}
Demographic models inferred from genetic data play several important roles in population genetics.
First, they complement archeological evidence in understanding prehistorical events (such as the number and timing of major continental migrations) which have left no written record~\cite{bib:Mellars2006,bib:Goebel2008}.
Second, they facilitate the search for genetic regions that have been targets of non-neutral forces, such as recent natural selection, by guiding our expectations as to how much sequence and haplotype variation one expects to see in a given genomic region (and, more importantly, the variance around these expectations)~\cite{bib:Nielsen2007}.
Finally, existing demographic models can guide sampling design for subsequent population or medical genetic studies.

Given their many uses, it is not surprising that many studies have inferred demographic models for populations of humans and other species~\cite{bib:Adams2004, bib:Marth2004, bib:Voight2005, bib:Hey2005, bib:Schaffner2005, bib:Becquet2007, bib:Caicedo2007, bib:Keinan2007, bib:Garrigan2007, bib:Mulligan2008, bib:Kitchen2008, bib:Cox2008}.

The process of inferring a demographic model consistent with a particular data set typically involves exploring a large parameter space by simulating the model many times, often using coalescent-theory based Monte Carlo approaches.  For computational reasons, many of the demographic inference procedures developed thus far have focused on single population models or models with multiple populations but no subsequent migration after subpopulations split (i.e., \cite{bib:Adams2004,bib:Marth2004, bib:Drummond2005, bib:Voight2005, bib:Hernandez2007}, but also see \cite{bib:Caicedo2007,bib:Nielsen2009}).  Methods that do consider multiple populations with migration often assume independent non-recombining regions~\cite{bib:Hey2004,bib:Hey2005} and do not often scale to genomic size data sets.  Approaches for jointly considering recombination and migration often use a restricted set  of summary statistics~\cite{bib:Becquet2007} of the data, which limits their statistical power. Finally, complex demographic inferences that make use of many summary statistics are often very computationally intensive~\cite{bib:Schaffner2005,bib:Caicedo2007,bib:Nielsen2009}, which precludes thorough investigation of their statistical properties.  

Here, we develop and apply a computationally efficient diffusion-based approach to the problem of demographic inference, based on the multi-population allele frequency spectrum (AFS) (i.e., the joint distribution of allele frequencies across SNPs)~\cite{bib:Sawyer1992,bib:Bustamante2001,bib:Hernandez2007,bib:Caicedo2007,bib:Nielsen2009}.
Given a genetic region sequenced in multiple individuals from each of $P$ populations, the resulting AFS is a $P$-dimensional matrix.
Each entry of this matrix records the number of diallelic genetic polymorphisms in which the derived allele was found in the corresponding number of samples from each population.
For example, if diploid individuals from two populations were sequenced, with 10 individuals from population 1 and 5 from population 2, the AFS would be a 21-by-11 matrix (indexed from 0).
The [2,0] entry would record the number of polymorphisms for which the derived allele was seen twice in population 1 but never seen in population 2, while the [20,5] entry would record polymorphisms for which the derived allele was homozygous in all individuals from population 1 and seen 5 times in population 2.
If all polymorphic sites possess only two alleles and can be considered independent, the AFS is a complete summary of the data. Many of the statistics commonly used for population genetic inference, such as $F_{ST}$ and Tajima's~$D$, are summaries of the AFS (see \cite{bib:Wakeley2008,bib:Nielsen2009}).

Efficient techniques exist for simulating the AFS of a single population~\cite{bib:Adams2004,bib:Marth2004,bib:Williamson2005}.  The joint AFS between two populations has been used by several recent studies~\cite{bib:Hernandez2007a,bib:Caicedo2007,bib:Keinan2007,bib:Nielsen2009}, but these have all relied upon very computationally intensive coalescent simulations.
Here we approximate the joint multi-population AFS by numerical solution of a diffusion equation, and our implementation supports up to three simultaneous populations.
Because the diffusion approach neglects linkage, our comparison with the data is through a composite likelihood function. 
Such likelihoods are consistent estimators under a wide range of population genetic scenarios for selectively-neutral data, but do not correctly capture variances~\cite{bib:Wiuf2006}. (Lower recombination induces higher linkage and higher variance among the expected entries of the AFS.) As we demonstrate below, the efficiency of our diffusion approach enables both conventional and parametric bootstrap resampling of the data, allowing us to accurately estimate confidence intervals for parameter values and critical values for hypothesis tests~\cite{bib:Zhu2005}, accounting for any degree of linkage found in the data.
This bootstrap procedure overcomes the traditional concerns with composite likelihood as a philosophy for inference in population genetics

To demonstrate the utility of our approach, we apply our method to two epochs in human history, using single nucleotide polymorphism (SNP) data from the Environmental Genome Project (EGP)~\cite{bib:Livingston2004}, the largest public database of human resequencing data.
We first study the expansion of humans out of Africa, jointly modeling the history of African, European, and East Asian populations.
We then study the settlement of the New World, jointly modeling European, East Asian, and admixed Mexican populations. In both cases, we quantify the uncertainty of our parameter inferences and test hypotheses about migration (bootstrapping to account for linkage).
In particular, we infer an earlier divergence between African and Eurasian populations than previous studies, because our inferences account for the substantial migration between these populations.
Our methods also find no evidence for multiple migrations between East Asia and the New World.
While similarly complex models for human continental populations have been studied~\cite{bib:Schaffner2005}, to our knowledge, our analysis is the first in which the full joint AFS is used for inference and in which uncertainty and goodness-of-fit have been quantified.

An important advantage of the diffusion approach is the ease with which selection can be incorporated.
As an illustrative application, we also predict the distribution of protein-coding variation between populations.  In agreement with the data, we find that less nonsynonymous variation is shared between populations than might be expected based only on patterns of shared noncoding variation. 

While no model can capture the full complexity of any species' genetic history, the models presented refine our understanding of the expansion of humanity across the globe. None of the methodology is specific to humans, and we expect our method will find wide application to demographic inference of other species.

\onecolumngrid

\section{Methods}

\subsection{Diffusion approximation}

To efficiently simulate the AFS, we adopt a diffusion approach. 
Such approaches have a long and distinguished history in population genetics, dating back to R.\ A.\ Fischer~\cite{bib:Fischer1922,bib:Kimura1964,bib:Ewens2000}.
The diffusion approach is a continuous approximation to the population genetics of a discrete number of individuals evolving in discrete generations.
An important underlying assumption is that per-generation changes in allele frequency are small.
Consequently, the diffusion approximation applies when the effective population size $N$ is large and migration rates and selection coefficients are of order $1/N$.

If we have samples from $P$ populations, the numbers of sampled sequences from each population are $n_1, n_2, \dots, n_P$.
(For diploids, $n_1$ is twice the number of individuals sampled from population 1.)
Entry $d_1, d_2, \dots, d_P$ of the AFS records the number of diallelic polymorphic sites at which the derived allele was found in $d_1$ samples from population 1, $d_2$ from population 2, and so forth. (If ancestral alleles cannot be determined, then the ``folded'' AFS can be considered, in which entries correspond to the frequency of the minor allele.)

\begin{figure*}
\centering
\includegraphics[width=\textwidth]{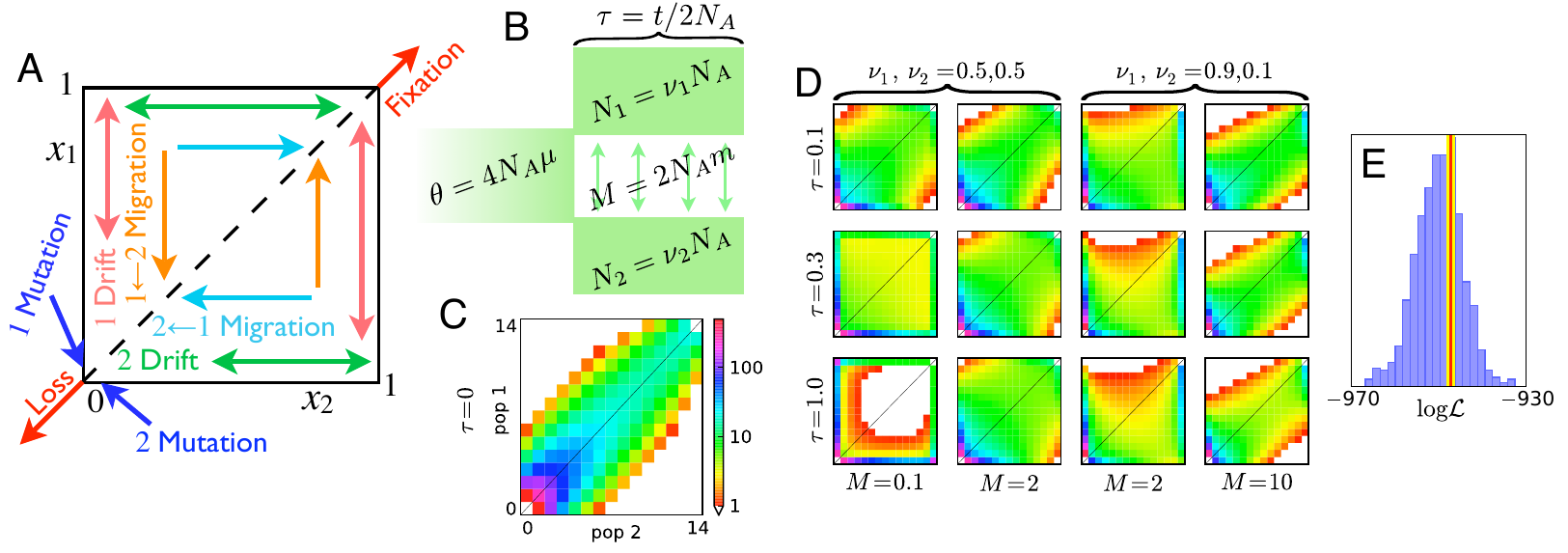}
\caption{
{\bf Frequency spectrum gallery.}
A) Qualitative effects of modeled neutral genetic forces on $\phi(x_1, x_2, t)$, the density of alleles at relative frequencies $x_1$ and $x_2$ in populations 1 and 2.
B) For the spectra shown, an equilibrium population of effective size $N_A$ diverges into two populations $2N_A \tau$ generations ago.
Populations 1 and 2 have effective sizes $\nu_1 N_A$ and $\nu_2 N_A$, respectively. Migration is symmetric at $m = M/(2 N_A)$ per generation, and $\theta = 1000$.
C) The AFS at $\tau = 0$. Each entry is colored by the logarithm of the number of sites in it, according to the scale shown.
D) The AFS at various times for various demographic parameters, on the same scale as B.
E) Comparison between coalescent- and diffusion-based estimates of the likelihood $\mathcal{L}$ of data generated under the model A.
Coalescent-based estimates of the likelihood, each of which took approximately 7.0 seconds, are represented in the histogram.
The result from our diffusion approach, which took 2.0 seconds, is represented by the red line.
For accuracy comparison, the yellow line indicates the likelihood inferred from $10^6$ coalescent simulations.
}
\label{fig:gallery}
\end{figure*}

We model the evolution of $\phi(x_1, x_2, \dots, x_P, t)$, the density of derived mutations at relative frequencies $x_1, x_2, \dots, x_P$ in populations $1, 2, \dots, P$ at time $t$.
(All $x$ run from 0 to 1.)
Given an infinitely-many-sites mutational model~\cite{bib:Watterson1975} and Wright-Fisher reproduction in each generation, the dynamics of $\phi$ for an arbitrary finite number of populations $P$ are governed by a linear diffusion equation:
\begin{align}\label{eqn:main}
\frac{\partial }{\partial \tau} \phi =
\frac{1}{2} \sum_{i = 1,2,\dots,P} \frac{\partial^2}{\partial^2 x_i} \, \frac{x_i (1-x_i)}{\nu_i} \, \phi
 -\sum_{i = 1,2,\dots,P} \frac{\partial}{\partial x_i} \bigg( \gamma_i x_i (1-x_i) + \sum_{j=1,2,\dots,P} \, M_{i \leftarrow j} (x_j - x_i)\bigg) \phi.
\end{align}
The first term models genetic drift, and the second term models selection and migration.
Fig~\ref{fig:gallery}A illustrates the effects of different evolutionary forces on components of $\phi$.
Time is in units of $\tau = t/(2 N_{ref})$, where $t$ is the time in generations and $N_{ref}$ is a reference effective population size.  
The relative effective size of population 1 is $\nu_1 = N_1/N_{ref}$.
The scaled migration rate is $M_{1 \leftarrow 2} = 2 N_{ref} m_{1 \leftarrow 2}$, where $m_{1 \leftarrow 2}$ is the proportion of chromosomes per generation in population 1 that are new migrants from population 2.
(Thus migration is assumed to be conservative~\cite{bib:Nagylaki1980}).
Finally, the scaled selection coefficient is $\gamma_1 = 2 N_{ref} s_1$, where $s_1$ is the relative selective advantage or disadvantage of variants in population 1.
Boundary conditions are no-flux except at two corners of the domain, where all population frequencies are 0 or 1; these are absorbing points corresponding to allele loss or fixation.
Because the diffusion equation is linear, we can solve simultaneously for the evolution of all polymorphism by continually injecting $\phi$ density at low frequency in each population (at a rate proportional to the total mutation flux $\theta$), corresponding to novel mutations.

Changes in population size and migration alter the parameters in Eqn~\ref{eqn:main}, while population splits and mergers alter the dimensionality of $\phi$. 
For example, if new population 3 is admixed with a proportion $f$ from population 1 and $1-f$ from population 2 then
\begin{equation}\label{eqn:split}
\phi(x_1, x_2, x_3, t) = \phi(x_1, x_2, t) \, \delta\big(x_3 - [f x_1 + (1-f) x_2]\big),
\end{equation}
where $\delta$ denotes the Dirac delta function.
To remove population 2, $\phi$ is integrated over $x_2$:  $\phi(x_1, x_3, t) = \int_0^1 \phi(x_1, x_2, x_3, t) \,\textrm{d} x_2$.

Given $\phi$, the expected value of each entry of the AFS, $M[d_i, d_j, \dots,d_P]$, is found via a $P$-dimensional integral over all possible population allele frequencies of the probability of sampling $d_i, d_j\dots, d_P$ derived alleles times the density $\phi$ of sites with those population allele frequencies.
For SNP data obtained by resequencing, these probabilities are binomial, so
\begin{equation}
M[d_i, d_j, \dots, d_P] =
 \int_0^1 \cdots \int_0^1 \prod_{i = 1,2,\dots,P}  {n_i\choose d_i} x_i^{d_i} (1-x_i)^{n_i - d_i} \, \phi(x_1,x_2,\dots,x_P)\, \textrm{d}x_i.\label{eqn:to_M}
\end{equation}
In some cases of ascertained data~\cite{bib:Clark2005}, the resulting bias can be corrected by modifying the above equation~\cite{bib:Keinan2007,bib:Nielsen2004}.

\subsection{Likelihood-based inference}
Let $\Theta$ correspond to the parameters of a demographic model we wish to estimate from the observed multi-population allele frequency spectrum, which we denote  $S[d_i, d_j, \dots]$.  Assuming no linkage between polymorphisms, each entry in the AFS is an independent Poisson variable~\cite{bib:Sawyer1992}, with mean $M[d_i, d_j, \dots]$ (which depends on $\Theta$).  We can, therefore, construct a likelihood function $\mathcal{L}(\Theta \mid S)$ using standard statistical theory:
\begin{equation}
\mathcal{L}(\Theta \mid S) = \prod_{i = 0 \dots P} \prod_{d_i = 0\dots n_i} \frac{e^{-M[d_i, d_j,\dots,d_P]} M[d_i, d_j,\dots,d_P]^{S[d_i, d_j,\dots,d_P]}}{S[d_i, d_j,\dots,d_P]!}.\label{eqn:likelihood}
\end{equation}

In words, our approach consists of calculating the expected allele frequency spectrum $M$ using a particular demographic model (and set of parameter values for that demographic model) using our diffusion approach.  We then maximize the similarity between $M$ and the observed AFS, $S$, over the parameter values that $\Theta$ can take on. Competing demographic models can be chosen from using standard statistical theory such as the nested likelihood ratio test or Information Criteria such as the Akaike or Bayesian Information Criteria.

For linked polymorphisms, $\mathcal{L}$ is a composite likelihood.
Such likelihoods are consistent estimators under a wide range of neutral population genetic scenarios~\cite{bib:Wiuf2006}, but simulations incorporating linkage are necessary to estimate variances and define critical values for hypothesis testing and model selection.
In our applications, we estimate variances using simulations from the coalescent simulator \emph{ms}~\cite{bib:Hudson2002}.

\subsection{Numerics}
Solving the multi-population diffusion equation is substantially more demanding than the single-population case~\cite{bib:Williamson2005}. This is primarily because the boundary conditions are more complex, and the numerical grid of population frequencies $x$ must be much coarser to be computationally tractable, because it is of $P$ dimensions.
For example, a previous single-population study~\cite{bib:Williamson2005} used a uniform $x$ grid of order $10^4$ values between 0 and 1.
Extending this grid to a three-population simulation would require an infeasible array of size $10^{12}$.
Instead, we use a nonuniform grid and extrapolation to enable accurate computation using of order 100 values along each dimension, for a final array size of order $10^6$.

We solve the diffusion equation on a regular nonuniform grid, using a finite difference scheme~\cite{bib:Press2007} inspired by the method of Chang and Cooper~\cite{bib:Chang1970} (supporting information).
Mutations in population 1 arise at frequency $1/(2 N_1) = 1/(2 N_{ref} \nu_1)$.
The diffusion approximation applies when $N_{ref} \rightarrow \infty$, but the minimum frequency in our numerical simulation is that of the first grid point, denoted $\Delta$.
To overcome this, we extrapolate our results to an infinitely fine grid. 
We use a quadratic extrapolation on the logarithm of the AFS entry, modeling the bias introduced by the finite initial grid point $\Delta$ as 
\begin{equation}
\log M_\text{calc}(\Delta) = \log M_\infty + a  \Delta + b \Delta^2.\label{eqn:extrap}
\end{equation}
Here $M_\text{calc}(\Delta)$ is an AFS element calculated at grid size $\Delta$ and $M_\infty$ is the extrapolated value.
Given three evaluations at different grid sizes $\Delta$, we solve for $M_\infty$ and use this value when calculating likelihoods.
This vastly increases both the speed and accuracy of our calculation (supporting information).
While higher-order extrapolations may improve accuracy in some cases, they may also be more sensitive to numerical noise.
Our empirical experience is that a quadratic approximation provides a good compromise between accuracy, efficiency, and robustness.

The computational cost for a single likelihood evaluation scales as $G^{P+1}$ where $G$ is the number of grid points used.
In our experience, for stability and accuracy $G$ should somewhat larger than the largest population sample size.
Although our theoretical framework extends to an arbitrary number of populations, the exponential scaling of computation with $P$ limits our current applications to three simultaneous populations.
Importantly, our likelihood calculation is deterministic and numerically smooth, so numerical derivatives can be used in optimization.
We use the the quasi-Newton BFGS method~\cite{bib:Press2007}, which converges in order $N_p^2$ steps, where $N_p$ is the number of free parameters.

Our implementation of these methods, \dadi, is written in cross-platform Python and C, making use of the NumPy~\cite{bib:Oliphant2006}, Scipy~\cite{bib:Oliphant2007}, and Matplotlib libraries~\cite{bib:Hunter2007}.
It is distributed under the open-source BSD license.
All calculations herein were performed with \dadi\ version 1.1.0.

We estimated parameter uncertainties by both conventional bootstrap (fitting data sets resampled over loci) and parametric bootstrap (fitting simulated data sets).
To generate simulated data we used the coalescent program \emph{ms}~\cite{bib:Hudson2002}, a region-specific recombination rate, and the detailed EGP sequencing strategy (supporting information).

The confidence intervals reported in Tables~\ref{tbl:YRIfit} and~\ref{tbl:MXLfit} derive from a normal approximation to the bootstrap results.
For the conventional bootstrap, confidence intervals were calculated as:
$\overline{\theta^*}  \pm 1.96  \, \sigma(\theta^*)$.
For the parametric bootstrap, biased-corrected intervals were calculated as:
$\hat\theta - (\overline{\theta^*} - \hat\theta) \pm 1.96  \, \sigma(\theta^*)$.
The maximum-likelihood value is denoted $\hat \theta$, while $\overline{\theta^*}$ and $\sigma(\theta^*)$ denote the mean and standard deviation of the bootstrap results.
Aside from the growth rates $r$, all our model parameters are positive by definition, so in those cases we used their logarithms when calculating confidence intervals.

Pearson's $\chi^2$ goodness-of-fit test was performed using all $21^3 - 2$ = 9259 bins in the AFS. Results are similar if we restrict our analysis to entries in which the expected value is $>1$ or $>5$.

\subsection{Data}
We used the National Institute of Environmental Health Science's Environmental Genome Project SNPs database~\cite{bib:EGP},
which results from direct Sanger resequencing of environmental response genes in several populations.
We considered all diallelic SNPs in 5.01 Mb of sequence from noncoding regions of 219 autosomal genes (supporting information).
These data have been the subject of many publications, including~\cite{bib:Akey2004, bib:Livingston2004, bib:Williamson2005, bib:Hernandez2007}.
As an assessment of quality, additional high-coverage short-read sequencing has recently been performed across 8 samples in this data set. Over 26,000 sites, the SNP concordance between this next-generation sequencing and the original Sanger sequencing averages 99.5\% (D.\ Nickerson, personal communication).
Given the high quality of this data set, we do not incorporate sequencing error into our modeling. 
We believe such correction will be essential in future applications to less accurate short-read sequencing data, as inference based on the frequency spectrum is sensitive to rare alleles.

To estimate the ancestral allele, we aligned to the panTro2 build of the chimp genome~\cite{bib:Chimpanzee2005}.
Like other methods based on the unfolded AFS, our analysis is sensitive to errors in identifying the ancestral allele.
We statistically corrected the AFS for ancestral misidentification~\cite{bib:Hernandez2007}, using a context-dependent substitution model~\cite{bib:Hwang2004}.
This procedure has been shown to perform better than aligning to multiple species~\cite{bib:Hernandez2007}.
To account for missing data and ease qualitative comparisons between populations, we projected all spectra down to 20 samples per population~\cite{bib:Marth2004} (supporting information).

The human-chimp divergence in the data is 1.13\%.
We assumed a divergence time of 6 My~\cite{bib:Kumar2005} and a generation time of 25 years.
This yielded an estimated neutral mutation rate of $\mu = 2.35 \times 10^{-8}$ per site per generation, which is comparable to direct estimates~\cite{bib:Kondrashov2002}.
There is some controversy as to the appropriate generation time to assume in human population genetic studies~\cite{bib:Fenner2005,bib:Tremblay2000}.
In particular, the human generation time may differ between cultures and may have changed during our biological and cultural evolution.
The bootstrap uncertainties reported in Tables~\ref{tbl:YRIfit} and~\ref{tbl:MXLfit} do not include systematic uncertainties in the human-chimp divergence or generation times.
The generation time, however, formally cancels when converting between genetic and chronological times.

\subsection{Nonsynonymous polymorphism}
In our prediction of the distribution of nonsynonymous polymorphism, the distribution of selective effects assumed was a negative-gamma distribution with shape parameter $\alpha = 0.184$ and scale $\beta = 8200$~\cite{bib:Boyko2008}. The AFS was calculated by trapezoid-rule integration over this distribution, using 201 evaluations logarithmically spaced over $\gamma = [-300, -10^{-6}]$. All demographic parameters, including the scaled mutation rate $\theta$, were set to the maximum-likelihood values from our Out of Africa analysis.

\vspace{0.25in}
\hrulefill
\vspace{0.25in}

\twocolumngrid

\section{Results}

First, we explored how various demographic forces affect the AFS, building intuition for our subsequent applications to real data.
We then compared the performance of diffusion versus coalescent methods for evaluating the AFS, finding that the diffusion approach is substantially faster.
We then applied our diffusion approach to infer parameters for plausible demographic models for the history of continental human populations. We first considered the expansion of humans out of Africa and then the settlement of the New World. In these applications, we inferred the maximum composite-likelihood parameters of our models using diffusion fits to the real data.  To account for linkage in estimating variances and critical values for hypothesis tests, we then repeatedly fit both conventional and parametric bootstrap data sets.
Finally, in an application incorporating selection, we predicted the distribution of nonsynonymous variation between populations in our Out of Africa model, finding good agreement with the available data.

\subsection{Demographic effects on the AFS}

In Fig~\ref{fig:gallery}, we provide examples of the AFS under different demographic scenarios.
Fig~\ref{fig:gallery}B illustrates the isolation-with-migration model for which the spectra are calculated.
The expected spectrum at zero divergence time is shown in Fig~\ref{fig:gallery}C.
Fig~\ref{fig:gallery}D shows the expected spectrum at various divergence times under various demographic scenarios.
Qualitatively, correlation between population allele frequencies declines with increasing divergence time, depopulating the diagonal of the AFS.
On the other hand, migration prolongs and sustains correlation.
Less obviously, AFS entries corresponding to shared low-frequency alleles distinguish between increased migration and reduced divergence time (supporting information).
Additionally, differences in genetic drift between populations with different effective sizes result in asymmetries in the AFS.
These qualitative features of the AFS are also evident in human data; detailed modeling allows us to quantify our inference regarding the type, timing, and strength of demographic events that are consistent with the data.

\subsection{Computational performance}

The computer program implementing our method is named \dadi\ (Diffusion Approximations for Demographic Inference). It is open-source and freely available at {\tt http://dadi.googlecode.com}.

Fig~\ref{fig:gallery}E compares \dadi\ with a coalescent approach to evaluating the likelihood of frequency spectrum data.
The coalescent simulator \emph{ms}~\cite{bib:Hudson2002} was used to generate a simulated data set from the model in Fig~\ref{fig:gallery}B, with
parameters $\nu_1=0.9$, $\nu_2 = 0.1$, $M=2$, $\tau = 2$, $\theta = 1000$, scaled total recombination rate $\rho=1000$, and 20 samples per population.
Coalescent-based estimates of the expected AFS were generated by averaging $10^5$ \emph{ms} simulations, each run with $\theta = 1$ and $\rho = 0$.
These estimates were scaled to $\theta = 1000$ for comparison with the simulated data set. (This procedure is substantially faster than simulating with larger $\theta$ and $\rho$.)
Each estimate took approximately 7.2 seconds of computation.
The histogram in Fig~\ref{fig:gallery}E shows the resulting distribution of estimated likelihoods of the data.
Shown by the red line in Fig~\ref{fig:gallery}E is the result from our diffusion approach (with grid sizes $G = \{40,50,60\}$), which took approximately 2.0 seconds of computation.
The yellow line is the likelihood from $10^8$ coalescent simulations, illustrating the high accuracy of our diffusion approach.
(Note that the coalescent approach we consider here is not necessarily optimal. We are, however, unaware of any such approach that is competitive in computational speed with the diffusion method.)

The computational advantage of the diffusion method is even larger when placed in the context of parameter optimization.  Unlike the coalescent approach, there is no simulation variance, so efficient derivative-based optimization methods can be used. 
As examples, consider our applications to human data, which involve 20 samples per population.
On a modern workstation, fitting a single-population three-parameter model took roughly a minute, while fitting a two-population six-parameter model took roughly 10 minutes.
The fits of three-population models with roughly a dozen parameters typically took a few hours to converge from a reasonable initial parameter set.
This speed allows us to use extensive bootstrapping to estimate variances, overcoming the limitations of composite likelihood.

\subsection{Expansion out of Africa}

\begin{figure*}
\centering
\includegraphics{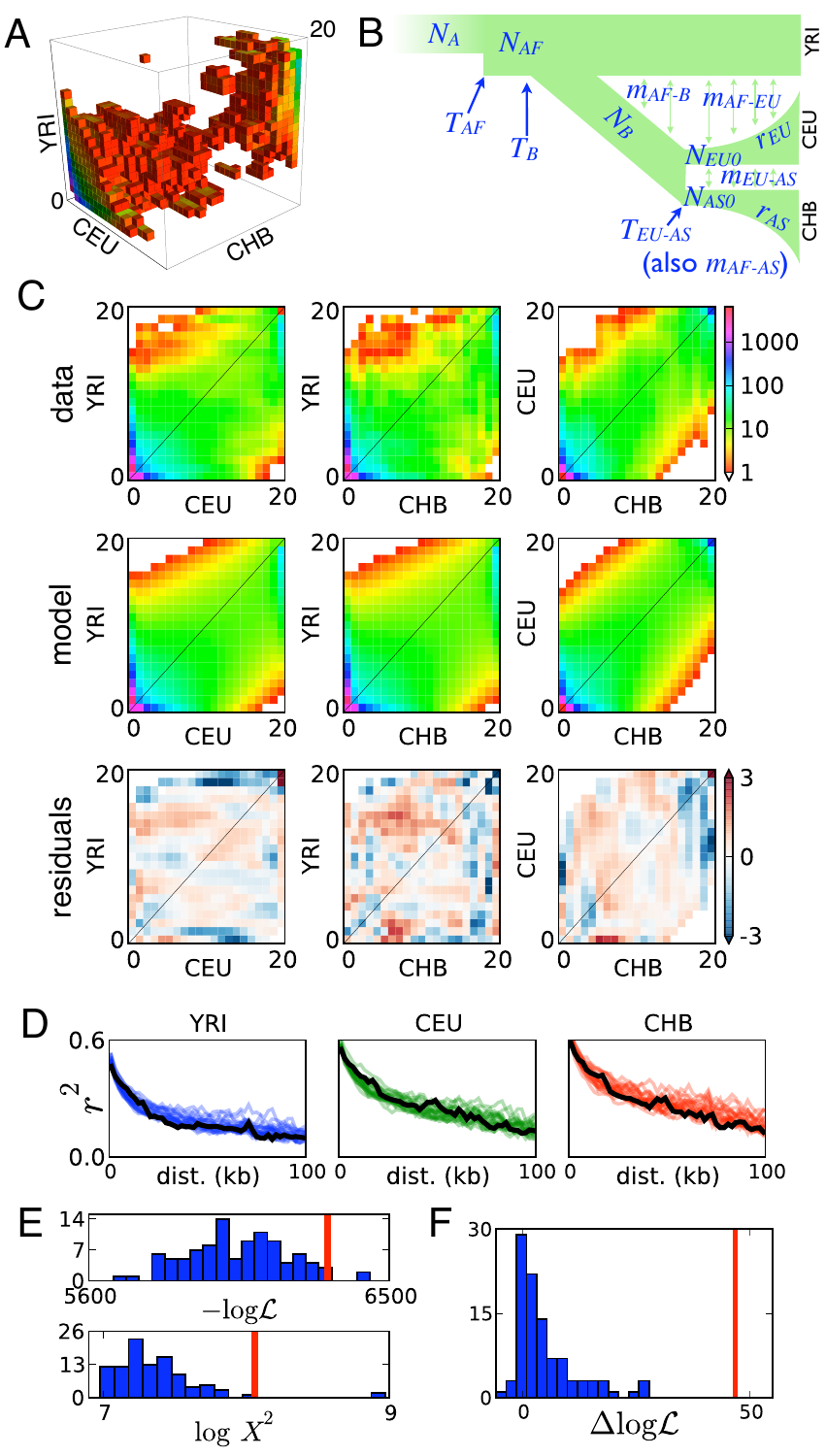}
\caption{
{\bf Out of Africa analysis.}
A) AFS for the YRI, CEU and CHB populations. The color scale is as in subfigure C.
B) Illustration of the model we fit, with the 14 free parameters labeled.
C) Marginal spectra for each pair of populations. The top row is the data, and the second is the maximum-likelihood model. The third row shows the Anscombe residuals~\cite{bib:Pierce1986} between model and data. Red or blue residuals indicate that the model predicts too many or too few alleles in a given cell, respectively.
D) The observed decay of linkage disequilibrium (black lines) is qualitatively well-matched by our simulated data sets (colored lines).
E) Goodness-of-fit tests based on the likelihood $\mathcal{L}$ and Pearson's $X^2$ statistic both indicate that our model is a reasonable, though incomplete description of the data. In both plots, the red line results from fitting the real data and the histogram from fits to simulated data. Poorer fits lie to the right (lower $\mathcal{L}$ and higher $X^2$).
F) The improvement in likelihood from including contemporary migration in the real data fit (red line) is much greater than expected from fits to simulated data generated without contemporary migration (histogram).
This indicates that the data contain a strong signal of contemporary migration.
}
\label{fig:YRIfit}
\end{figure*}

Our analysis of human expansion out of Africa used data from three HapMap populations: 12 Yoruba individuals from Ibadan, Nigeria (YRI); 22 CEPH Utah residents with ancestry from northern and western Europe (CEU); and 12 Han Chinese individuals sampled in Beijing, China (CHB).
Because approaches based on the frequency spectrum are sensitive to miscalling of the ancestral state, we statistically corrected for ancestral misidentification using an approach that accounts for a myriad of mutation and context-dependent biases (such as CpG effects)~\cite{bib:Hernandez2007}.
To ease qualitative comparison among populations and account for missing data, we projected the data down to 20 sampled chromosomes per population~\cite{bib:Marth2004}.
Because this data set is of very high quality ($>$99\% concordance of sequenced SNPs with next-generation sequencing of the same individuals to high coverage; see Materials and Methods), we do not explicitly correct for sequencing errors here.
We were left with 17,446 segregating diallelic single nucleotide polymorphisms (SNPs) from effectively 4.04 Mb of sequence.
Fig~\ref{fig:YRIfit}A shows the resulting AFS.  For ease of visualization, the top row of Fig~\ref{fig:YRIfit}C shows the two-population marginal spectra.

There are many possible three-population demographic models one could consider for these populations.
To develop a parsimonious yet realistic model, we first considered the marginal AFS for each population and each pair of populations.  Previous analyses found that the YRI spectrum is well-fit by a two-epoch model with ancient population growth~\cite{bib:Marth2004, bib:Hernandez2007}, and we found this as well (supporting information).  Previous analyses of the CEU and CHB populations found that both populations went through bottlenecks~\cite{bib:Marth2004, bib:Keinan2007} concurrent with divergence~\cite{bib:Keinan2007}.  Such models qualitatively fit the marginal CEU-CHB spectrum (supporting information).

\begin{table*}
\caption{
{\bf Out of Africa inferred parameters}}
\begin{minipage}{\textwidth}
\begin{tabular*}{\hsize}{@{\extracolsep{\fill}}cccc}
& &  conventional & parametric bootstrap\\
& maximum &  bootstrap 95\% & bias-corrected 95\%\\
parameter\footnote{See Fig~\ref{fig:YRIfit}B for model schematic. Growth rates $r$ and migration rates $m$ are per generation.} & likelihood  & confidence interval & confidence interval\\
\hline
$N_A$ &  7,300 & 4,400 -- 10,100 & 6,300 -- 9,200\\
$N_{AF}$ & 12,300 & 11,500 -- 13,900 & 11,100 -- 13,100\\
$N_B$ &  2,100 & 1,400 -- 2,900 & 1,700 -- 2,600\\
$N_{EU0}$ &  1,000 & 500 -- 1,900 & 500 -- 1,500\\
$r_{EU}$ (\%) &  0.40 & 0.15 -- 0.66 & 0.26 -- 0.57\\
$N_{AS0}$ &  510 & 310 -- 910 & 320 -- 750\\
$r_{AS}$ (\%) &  0.55 & 0.23 -- 0.88 & 0.32 -- 0.79\\
$m_{AF-B}$ ($\times 10^{-5}$) & 25 & 15 -- 34 & 19 -- 36\\
$m_{AF-EU}$ ($\times 10^{-5}$) & 3.0  & 2.0 -- 6.0 & 1.6 -- 7.6\\
$m_{AF-AS}$  ($\times 10^{-5}$) & 1.9 & 0.3 -- 10.4 & 0.7 -- 6.9\footnote{One low-migration outlier was removed for each of these estimations.}\\
$m_{EU-AS}$  ($\times 10^{-5}$) & 9.6 & 2.3 -- 17.4$^b$ & 5.7 -- 20.2\\
$T_{AF}$ (kya) & 220 &  100 -- 510 & 90 -- 410\\
$T_{B}$ (kya) & 140 & 40 -- 270 & 60 -- 310\\
$T_{EU-AS}$ (kya) & 21.2 & 17.2 -- 26.5 & 17.6 -- 23.9\\
\end{tabular*}
\end{minipage}
\label{tbl:YRIfit}
\end{table*}

Combining these demographic features yields the model illustrated in Fig~\ref{fig:YRIfit}B.
The maximum likelihood values for the 14 free parameters are reported in Table~\ref{tbl:YRIfit}.
Qualitatively, the resulting model reproduces the observed spectra well, as seen in the second and third rows of Fig~\ref{fig:YRIfit}C. 
(The correlation between adjacent residuals is due in part to our projection of the data down from a larger sample size (supporting information).)
Allowing for asymmetric gene flow yielded very little improvement in fit, as did allowing for growth in the Eurasian ancestral population or allowing the CEU and CHB bottleneck and divergence times to differ (data not shown).

Our composite likelihood function assumes that polymorphic sites are independent.
Because it thus overestimates the number of effective independent data points, confidence intervals calculated directly from the composite likelihood function will be too liberal. To control for linkage, we performed both conventional and parametric bootstraps.  Because our sequenced genes are typically well separated, they can be treated as independent, and our conventional bootstrap resampled from the 219 sequenced loci. For the parametric bootstrap, simulated data sets that incorporate linkage and the EGP's sequencing strategy were generated with \emph{ms}~\cite{bib:Hudson2002}. 

Table~\ref{tbl:YRIfit} reports parameter 95\% confidence intervals from both the conventional and bias-corrected parametric bootstraps.
The parametric bootstraps yield slightly smaller confidence intervals than the conventional bootstrap, suggesting that some variability in the data has not been accounted for by our simulations.
This variability may involve small varied selective forces on the sequenced regions, or slight relatedness between sampled individuals.
The parametric bootstrap results additionally show that our method possesses very little bias in parameter inference (supporting information).

As seen in Table~\ref{tbl:YRIfit}, the times for growth in the African ancestral population and divergence of the Eurasian ancestral population ($T_{AF}$ and $T_{B}$) have particularly wide confidence intervals, likely a consequence of the high inferred migration rate $m_{AF-B}$ between the African and Eurasian ancestral populations.
$T_{AF}$ shows high correlation with the ancestral population size $N_A$, while $T_B$ shows no strong linear correlation with any other single parameter (supporting information).
We found that 92 out of our 100 conventional bootstrap fits yield $N_{AS0} < N_{EU0}$, supporting the contention that the CHB population suffered a more severe bottleneck than the CEU population~\cite{bib:Keinan2007}.

We used several metrics to assess our model's goodness-of-fit, in additional to visual inspection of the residuals seen in Fig~\ref{fig:YRIfit}C.
Fig~\ref{fig:YRIfit}D compares the decay of linkage disequilibrium (LD) in the data and in the parametric bootstrap simulations.
The agreement seen is notable because our demographic inference used no LD information in building and fitting the model.
This LD comparison thus serves as independent validation of both our model and bootstrap simulations
We also asked whether the likelihood $\mathcal{L}$ found in the real data fit is atypical of fits to simulated data.
Out of fits to 100 simulated data sets, 2 produced a smaller likelihood (worse fit) than the real data fit (Fig~\ref{fig:YRIfit}E), yielding a p-value of $\approx$0.02.
One can craft examples in which a likelihood-based goodness-of-fit test fails to exclude very poor models~\cite{bib:Heinrich2001}.
Thus we also applied Pearson's $\chi^2$ goodness-of-fit test, a more robust and standard method for data that is in Poisson-distributed bins, such as the AFS~\cite{bib:Press2007}. 
In our case, we must use our parametric bootstraps to assess the significance of the sum-of-squared-residuals test statistic $X^2$, because many entries in the AFS are small and because they are not strictly independent.
Fig~\ref{fig:YRIfit}E shows the bootstrap-derived empirical distribution of $X^2$.
Two of the bootstraps yielded a larger $X^2$ (worse fit) than the real data fit, giving a p-value of $\approx$0.02, identical to that from the likelihood-based test.
(The two simulations that yield a higher $X^2$ than the real fit are not the same two that yield a lower $\mathcal{L}$, suggesting that these tests are somewhat independent.)
In some cases specific frequency classes of SNPs, such as rare alleles, may be of particular interest.
In the supporting information, we provide comparisons of the joint distribution of rare alleles seen in the data with that from our simulations.
These comparisons indicate that our model also reproduces well this interesting region of the frequency spectrum.
Finally, in Fig~\ref{fig:coding} we compare the model and data using larger bins of SNPs specific to specific populations or segregating at high or low frequency.
In all cases the model agrees within the uncertainty of the bootstrapped data.
Taken together, these tests suggest that our model provides a reasonable, though not complete, explanation of the data, lending credence to our demographic estimates.

The inferred contemporary migration parameters ($m_{AF-EU}$, $m_{AF-AS}$ and $m_{EU-AS}$) are small, raising the question as to whether they are statistically distinguishable from zero.
Figure~\ref{fig:YRIfit}F shows that the improvement in fit to the real data upon adding contemporary migration to the model is much larger than would be expected if there were no such migration, implying that the contemporary migration we infer is highly statistically significant.
Omitting ancient migration ($m_{AF-B}$) reduced fit quality even more, indicating that the data also demand substantial ancient migration.

\subsection{Settling the New World}

\begin{figure}[ht!]
\centering
\includegraphics[width=\columnwidth]{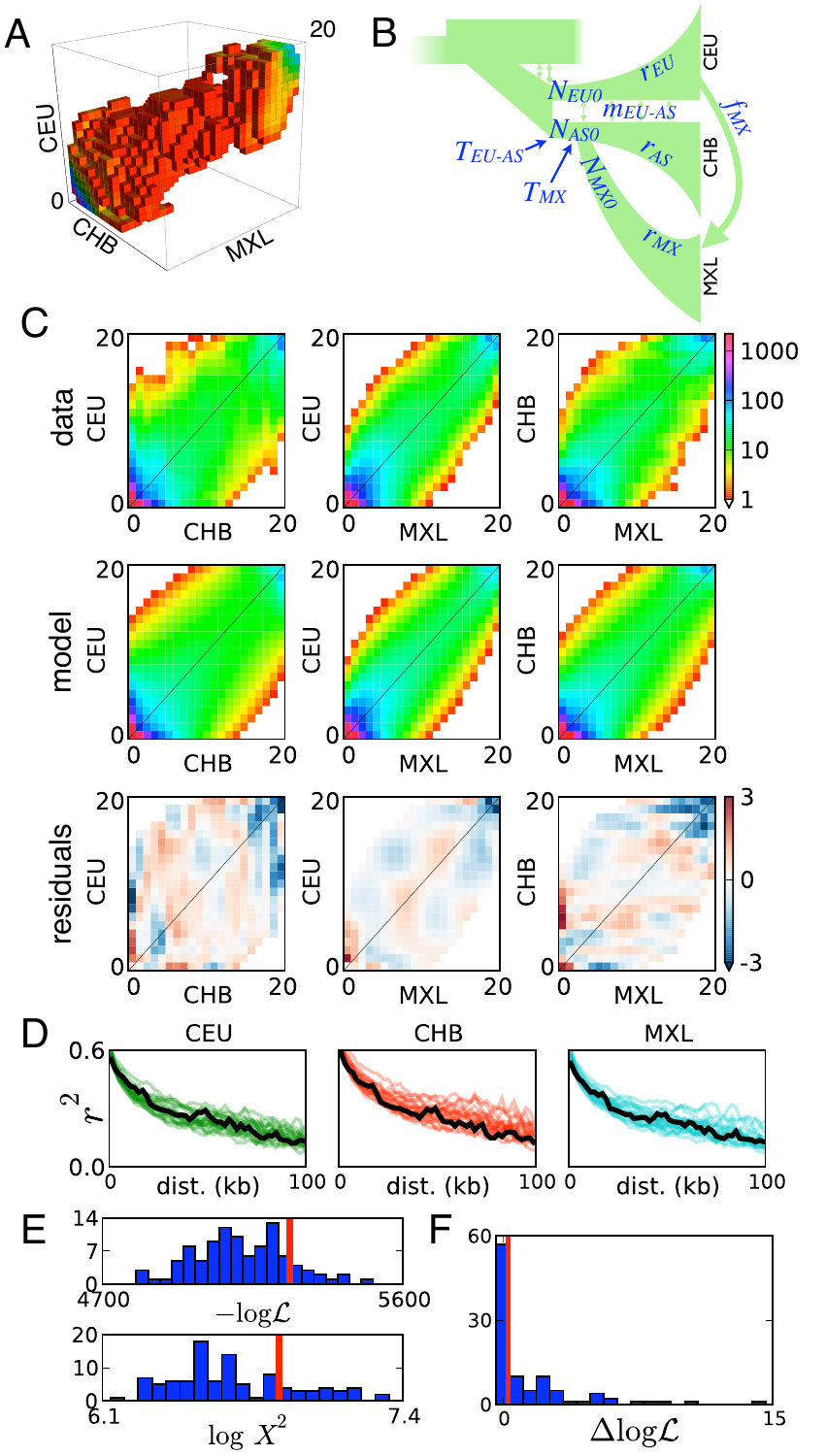}
\caption{
{\bf Settlement of the New World analysis.}
As in Fig~\ref{fig:YRIfit}, A) is the data, B) is a schematic of the model we fit, C) compares the data and model AFS, and D) compares LD.
E) The fit of our model to the real data is not atypical of fits to simulated data.
F) The improvement in real data fit upon including CHB-MXL migration (red line) is very typical of the improvement in fits to simulated data without CHB-MXL migration. Thus we have no evidence for CHB-MXL migration after divergence.
}
\label{fig:MXLfit}
\end{figure}

To study the settlement of the Americas, we used the previously considered 22 CEU and 12 CHB individuals plus an additional 22 individuals of Mexican descent sampled in Los Angeles (MXL).
Data were processed as in our Out of Africa analysis, yielding 13,290 segregating SNPs from effectively 4.22 Mb of sequence.
Fig~\ref{fig:MXLfit}A shows the resulting AFS, while Fig~\ref{fig:MXLfit}C shows the marginal spectra.

A model in which the CEU and CHB diverge from an equilibrium population did not reproduce the AFS well (supporting information).
Interestingly, a model allowing a prior size change in the ancestral population better fit the AFS but very poorly fit the observed LD decay (supporting information).
Thus, reproducing the AFS does not guarantee reproduction of LD, at least given a historically unrealistic model.
To develop a more realistic model, we endeavored to include the effects of Eurasian divergence from and migration with the African population.
Computational limits precluded us from considering all 4 populations simultaneously, so we dropped the African population from the simulation upon MXL divergence (Fig~\ref{fig:MXLfit}B).

\begin{table*}
\caption{
{\bf Settlement of New World inferred parameters}}
\begin{minipage}{\textwidth}
\begin{tabular*}{\hsize}{@{\extracolsep{\fill}}cccc}
& &  conventional & parametric bootstrap\\
& maximum &  bootstrap 95\% & bias-corrected 95\%\\
parameter\footnote{See Fig~\ref{fig:MXLfit}B for model schematic. Growth rates $r$ and migration rates $m$ are per generation. $f_{MX}$ is the average European admixture proportion of the Mexican-Americans sampled.} & likelihood  & confidence interval & confidence interval\\
\hline
$N_{EU0}$ &  1,500 & 700 -- 2,100 & 900 -- 2,200\\
$r_{EU}$ (\%) &  0.23 & 0.08 -- 0.45 & 0.16 -- 0.34\\
$N_{AS0}$  &  590 & 320 -- 800 & 410 -- 790\\
$r_{AS}$ (\%) &  0.37 & 0.16 -- 0.60 & 0.24 -- 0.51\\
$N_{MX0}$ &  800 & 160 -- 1,800 & 140 -- 1,600\\
$r_{MX}$ (\%) &  0.50 & 0.14 -- 1.17 & 0.41 -- 0.98\\
$m_{EU-AS}$ ($\times 10^{-5}$) & 13.5 & 7.5 -- 32.2 & 9.9 -- 20.8\\
$T_{EU-AS}$ (kya) & 26.4 & 18.1 -- 43.1 & 21.7 -- 30.7\\
$T_{Mx}$ (kya) & 21.6 & 16.3 -- 26.9 & 18.6 -- 24.7\\
$f_{MX}$ (\%) & 48 & 42 -- 60 & 41 -- 55\\
\end{tabular*}
\end{minipage}
\label{tbl:MXLfit}
\end{table*}

Table~\ref{tbl:MXLfit} records the maximum-likelihood parameter values inferred for this model.
Because this fit did not include African data, we could not reliably infer demographic parameters involving the African population.
Thus, for this point estimate we fixed the Africa-related parameters $N_A$, $N_{AF}$, $N_B$, $m_{AF-B}$, $m_{AF-EU}$, $m_{AF-AS}$, $T_{AF}$ and $T_{B}$ to their maximum-likelihood values from Table~\ref{tbl:YRIfit}.
Fig~\ref{fig:MXLfit}C compares the model and data spectra.
The residuals show little correlation, with the possible exception that the model may underestimate the number of high-frequency segregating alleles.

Parameter confidence intervals are reported in Table~\ref{tbl:MXLfit}.
To account for our uncertainty in those parameters derived from the Out of Africa fit, for each conventional bootstrap fit we used a set of Africa-related parameters randomly chosen from the sets yielded by our Out of Africa conventional bootstrap.
For the parametric bootstrap, we used the maximum-likelihood point estimates.
Again, we see that the conventional bootstrap confidence intervals are comparable to, although slightly wider than, the parametric bootstrap intervals.
Several parameters in this analysis have direct correspondence with our Out of Africa analysis.
Of particular note, the confidence intervals for the CEU-CHB divergence time $T_{EU-AS}$ overlap.

In assessing goodness of fit, Fig~\ref{fig:MXLfit}D shows that this model does indeed reproduce the observed pattern of LD decay.
Unlike in our Out of Africa analysis, however, here the LD decay was used to choose the form of the model (although not its parameter values), so this is not a completely independent assessment of fit.
Of our 100 parametric bootstrap fits, 13 yielded a worse likelihood than the real fit (Fig~\ref{fig:MXLfit}E), for a p-value of $\approx0.13$.
Applying Pearson's $\chi^2$ test, we find that 23 of 100 bootstrap fits yield a higher (worse) $X^2$ than the fit to the real data, for a p-value of $\approx0.23$, similar to that of the likelihood analysis.
Comparing distributions of rare alleles, our model typically reproduces the observed distribution well, although it may be somewhat overestimating the proportion of alleles that are rare or absent in the CHB population (supporting information).
In sum, our model appears to be a reasonable explanation of this data, somewhat better than in our Out of Africa analysis.

An essential feature of the Mexican-American individuals considered here is that they are typically admixed from Native American and European ancestors.
The $\approx$50\% average European admixture proportion we inferred for the MXL population is consistent with previous estimates for Los Angeles Latinos~\cite{bib:Price2007}.
We have no direct data from the Native American populations ancestral to MXL, but our model does account for their divergence from East Asia.
A model neglecting this divergence (by setting $T_{MX}$ to zero) fit the data substantially worse and yields an unrealistically high average European admixture proportion into MXL of 0.68.

Not only are Mexican-American individuals admixed, their admixture proportions also vary, and this subtlety is not directly accounted for in our analysis.
To assess its effect on our results, we first roughly estimated the ancestry proportion of each individual, using essentially a maximum-likelihood version~\cite{bib:Nielsen2009} of the algorithm used in \emph{structure}~\cite{bib:Pritchard2000} (supporting information).
(Methods based on ``admixture LD'', which identify breakpoints between regions of Native American and European ancestry, may be more powerful~\cite{bib:Patterson2004}. However, the strategy used by the EGP of sequencing widely spaced genes will resolve few of these breakpoints, limiting the applicability of these methods.) 
We then performed additional parametric bootstrap analyses, using simulations with a distribution of individual ancestry chosen to mimic that seen in the data and, to further test the method, with an extremely wide distribution.
These simulations showed that variation in individual ancestry does not bias our parameter inferences (supporting information).
Remarkably, it does not even change our statistical power.
This is evidenced by the fact that these bootstrap simulations yielded confidence intervals identical to our original simulations without variation in ancestry proportion (supporting information).
Nevertheless, future studies may profit by incorporating individual ancestry information~\cite{bib:Nielsen2009}, perhaps inferred from admixture LD.

Finally, our model allowed us to assess the role recurrent migration from Asia played in the settlement of the New World~\cite{bib:Goebel2008}.
When we added CHB-MXL migration to our model, we found that the maximum likelihood migration rate was $1.7 \times 10^{-5}$ per generation.
As shown in Fig~\ref{fig:MXLfit}F, the resulting improvement in likelihood is typical (p-value $\approx$0.45) of fits including CHB-MXL migration to data simulated without it.
Our data and analysis thus yielded no evidence of recurrent migration in the settlement of the New World.
Note, however, that this simple test does not necessarily rule out more complex scenarios, in which migration may vary over time.

\subsection{Nonsynomymous polymorphism}
Polymorphisms that change protein amino acid sequence are of medical interest because they are particularly likely to affect gene function~\cite{bib:Kryukov2009}.
Correspondingly, they are often subject to natural selection.
Diffusion approaches are particularly useful for studying such nonsynonymous polymorphism, because they easily incorporate selection.
Although the diffusion approximation assumes that sites are unlinked, nonsynonymous segregating sites are rare enough that this is often a reasonable approximation~\cite{bib:Boyko2008}.

\begin{figure}
\centering
\includegraphics[width=\columnwidth]{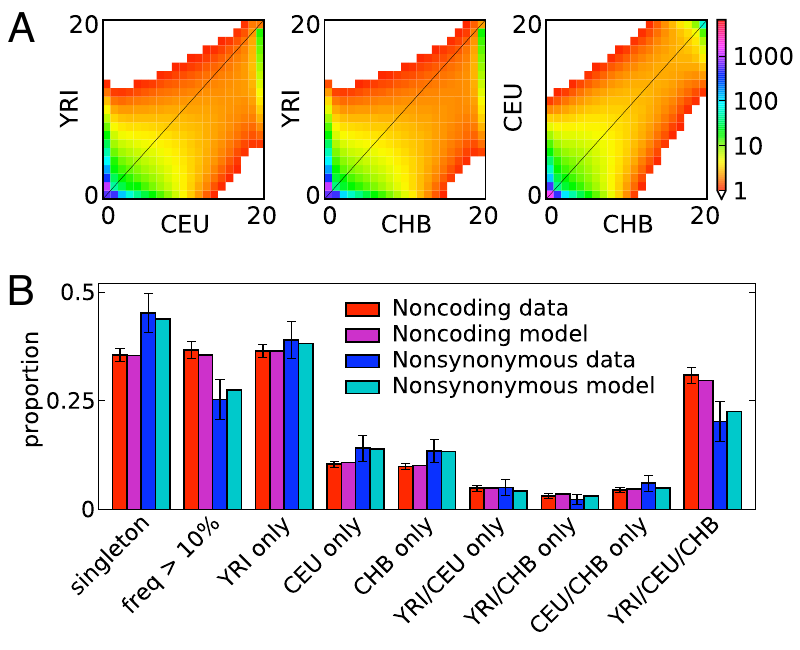}
\caption{
{\bf Distribution of nonsynonymous polymorphism.}
We simulated our maximum-likelihood Out of Africa demographic model with a distribution of selective effects previously inferred for nonsynonymous polymorphism~\cite{bib:Boyko2008}.
A) To enable direct comparison with the neutral AFS (Fig~\ref{fig:YRIfit}C), the scaled mutation rate $\theta$ was set identically, as is the color scale. As expected, selection dramatically reduces the amount of segregating polymorphism.
B) Shown are the proportions of variation found in various frequency classes. As expected, nonsynonymous variants typically have lower frequency. They also less likely to be shared between populations. Data error bars indicate 95\% bootstrap confidence intervals.\label{fig:coding}
}
\end{figure}

As an illustration, we used our Out of Africa demographic model to predict the distribution of such variation between continental populations.
To do so, we must specify a distribution for the selective effects of nonsynonymous mutations that enter the population.
For this we adopted a negative gamma distribution whose parameters were recently inferred~\cite{bib:Boyko2008}.
The resulting distribution of segregating variation is shown in Figure~\ref{fig:coding}A. 
(To ease comparison, we have assumed the same scaled mutation rate as in the neutral case of Fig~\ref{fig:YRIfit}C.)
As expected, selection sharply reduces the amount  of segregating polymorphism.
Figure~\ref{fig:coding}B shows the proportion of variants within various classes.
Also as expected, selection shifts nonsynonymous variation toward lower frequencies, raising the proportion of singletons and lowering the proportion at frequency greater than 10\%.
Less obviously, it also reduces the proportion of variation that is shared between populations.
In the neutral case, 43\% of polymorphism is predicted to be present in more than one population, while in the selected case only 35\% is. 
Thus genetic inferences from coding polymorphism may be less transferable between populations than might be expected from neutral patterns of allele sharing.

In the data considered here, there are about 400 nonsynomymous polymorphisms segregating in the three populations considered.
This is too few for a detailed goodness-of-fit test of our predicted distribution.
(Although see supporting information for a direct AFS comparison.)
Nevertheless, we observe that our predictions shown in Figure~\ref{fig:coding}B all lie within the bootstrap 95\% confidence intervals from the data.

\section{Discussion}
Our diffusion approximation to the joint allele frequency spectrum is a  powerful tool for population genetic inference. 
Although the diffusion approximation neglects linkage between sites, our method's computational efficiency allows us to use extensive bootstrap simulations to account for the effects of linkage.  (Let us reiterate that linkage does not affect the expected site-frequency spectrum of neutral sites, so our diffusion-based approach is estimating the same AFS that coalescent simulations are estimating, but in a small fraction of the time).
We applied our method to human expansion out of Africa and settlement of the New World, using public resequencing data from the Environment Genome Project.
The flexibility of the diffusion approach also allowed us to consider the distribution of non-neutral variation, which is difficult to address with other approaches.
Although no model can capture in detail the complete history of any population, the models presented here help refine our understanding of human expansion across the globe.

Our demographic results are broadly consistent with previous analyses of human populations.
In particular, single-population analyses have also inferred African population growth and European and Asian bottlenecks~\cite{bib:Adams2004,bib:Marth2004,bib:Voight2005}.
Also, the migration rates we infer are similar to those inferred by Schaffner et al.~\cite{bib:Schaffner2005} but somewhat smaller than those of Cox et al.~\cite{bib:Cox2008}.
On the other hand, Keinan et al.~\cite{bib:Keinan2007} inferred no significant migration between CEU and CHB.
Finally, our estimate of a New World founding effective population size in the hundreds is compatible other inferences~\cite{bib:Kitchen2008}.

Perhaps our most interesting demographic results are the inferred divergence times.
Other studies~\cite{bib:Keinan2007,bib:Garrigan2007} have estimated divergence times between Europeans and East Asians similar to the $\approx$23 kya we infer.
Interestingly, archeological evidence places humans in Europe much earlier ($\approx$40 kya)~\cite{bib:Mellars2006}.
Our inferred divergence time of $\approx$22 kya between East Asians and Mexican-Americans is somewhat older than the oldest well-accepted New World archeological evidence~\cite{bib:Goebel2008}.
The divergence we infer may reflect the settlement of Beringia, rather than the expansion into the New World proper~\cite{bib:Kitchen2008}.
Finally, the divergence time of $\approx$140 kya we infer between African and Eurasian populations is consistent with archeological evidence for modern humans in the Middle East $\approx$100 kya~\cite{bib:Mellars2006}, but it is much older than other inferences of $\approx$50 kya divergence from mitochondrial DNA~\cite{bib:Mellars2006}. 
This discrepancy may be explained by our inclusion of migration in the model.
Migration preserves correlation between population allele frequencies, so an observed correlation across the genome can be explained by either recent divergence without migration or ancient divergence with migration.
In fact, the African-Eurasian migration rate we infer of $\approx$$25\times 10^{-5}$ per generation is comparable to the $\approx$$100\times10^{-5}$ inferred from census records between modern continental Europe and Britain~\cite{bib:Weale2002}.

One difficulty in interpreting our divergence times is that the sampled populations may not best represent those in which historically important divergences occurred.
For example, the Yoruba are a West African population, so the divergence time we infer between Yoruba and Eurasian ancestral populations may correspond to divergence within Africa itself.
Future studies of more populations~\cite{bib:Li2008,bib:Jakobsson2008,bib:Wall2008} will help alleviate this difficulty.

Another difficulty is that the genic loci we study here may not be ideal for demographic inference.
Although we consider only noncoding sequence in fitting our historical model, selection on regulatory or linked coding sites may skew the AFS~\cite{bib:Braverman1995}.
In fact, the EGP data have been shown to differ in some ways (e.g. Tajima's~$D$) from intergenic regions~\cite{bib:Wall2008}.
Nevertheless, we use the EGP data because it is currently the largest public resource of noncoding human genetic variation, and we fit a neutral model because disentangling the small expected effects of selection on these sites from demographic effects will require additional data.
The rapidly declining cost of sequencing will give future studies access to many more loci that are likely to be less influenced by selection.
Importantly, the computational burden of our method is independent of the amount of sequence used to construct the AFS.
Additional loci will also increase power to discriminate between models and incorporate more detail.

The AFS encodes substantial demographic information.
It is has been shown, however, that an isolated population's AFS does not uniquely and unambiguously identify its demographic history~\cite{bib:Myers2008};
we expect a similar result to hold for multiple interacting populations.
Moreover, the AFS does not capture all the information in the data.
As illustrated by the alternative New World models we considered, patterns of linkage disequilibrium encode additional information.
Future studies may profit from coupling our efficient AFS simulation with methods that address other aspects of the data.

We have developed a powerful diffusion-based method for demographic inference from the joint allele frequency spectrum.
We applied our method to human expansion out of African and the settlement of the New World, developing models of human history that refine our knowledge and raise intriguing questions.
We also applied our method to predict the distribution of nonsynonymous variation across populations, and this prediction is consistent with the available data.
Our methods and the models inferred from it offer a foundation for studying the history and evolution of both our own species and others.

\begin{acknowledgments}
We thank Amit Indap for bioinformatics assistance and Jim Booth for statistical assistance. We also thank the NIEHS Program (ES-15478) for making their dataset so easily accessible. We had fruitful discussions with Adam Auton, Debbie Nickerson, Michael Hammer, Rasmus Nielsen, Nick Patterson, Molly Przeworski, Jeff Wall, and Carsten Wiuf. 
This research was supported by National Science Foundation grant PHY05-51164 and National Institutes of Health grants 1R01GM83606 and 2R01HG003229.
\end{acknowledgments}

\bibliography{popgen}

\end{document}